\documentclass[aps,prl,twocolumn,superscriptaddress]{revtex4-1}

\usepackage{amsfonts}%
\usepackage{amsmath}%
\usepackage{amssymb}%
\usepackage{graphicx}
\usepackage{color}

\usepackage{amstext}
\usepackage{rotating}
\usepackage{longtable}
\usepackage{url}
\usepackage{array}
\usepackage{multirow}

\newtheorem{theorem}{Theorem}

\newtheorem{definition}[theorem]{Definition}

\newcommand{\be}{\begin{equation}}
\newcommand{\ee}{\end{equation}}
\newcommand{\ba}{\begin{eqnarray}}
\newcommand{\ea}{\end{eqnarray}}
\newcommand{\av}[1]{\langle #1\rangle}

\definecolor{nred}{rgb}{0.7,0.2,0.2}
\definecolor{nblack}{rgb}{0,0,0}
\definecolor{nblue}{rgb}{0.2,0.2,0.7}
\definecolor{ngreen}{rgb}{0.2,0.6,0.2}

\begin{document}

\title{The definition of multipartite nonlocality}
\author{Jean-Daniel Bancal}
\affiliation{{Center for Quantum Technologies, National University of Singapore, Singapore}}
\author{Jonathan Barrett}
\affiliation{Department of Mathematics, Royal Holloway, University of London, Egham, Surrey TW20 0EX U. K.}
\author{Nicolas Gisin}
\affiliation{Group of Applied Physics, University of Geneva, Switzerland}
\author{Stefano Pironio}
\affiliation{Laboratoire d'Information Quantique, Universit\'e Libre de Bruxelles, Belgium}
\pacs{PACS number}

\begin{abstract}
In a multipartite setting, it is possible to distinguish quantum states that are genuinely $n$-way entangled from those that are separable with respect to some bipartition. Similarly, the nonlocal correlations that can arise from measurements on entangled states can be classified into those that are genuinely $n$-way nonlocal, and those that are local with respect to some bipartition. Svetlichny introduced an inequality intended as a test for genuine tripartite nonlocality. This work introduces two alternative definitions of $n$-way nonlocality, which we argue are better motivated both from the point of view of the study of nature, and from the point of view of quantum information theory. We show that these definitions are strictly weaker than Svetlichny's, and introduce a series of suitable Bell-type inequalities for the detection of $3$-way nonlocality. Numerical evidence suggests that all 3-way entangled pure quantum states can produce 3-way nonlocal correlations.
\end{abstract}
\maketitle

Consider two quantum systems, prepared in a joint quantum state $|\psi\rangle$ and located in separate regions of space. Suppose Alice measures one system, obtaining outcome $a$, and Bob the other, obtaining outcome $b$. The joint outcome probabilities can be written $P(ab|XY)$, where $X$ denotes Alice's measurement and $Y$ Bob's measurement. If the measurements are performed at spacelike separation, then Bell's condition of \emph{local causality} \cite{bellspeakable} implies that even if the particles have interacted in the past (or were produced together in the same source), they are now independent. Therefore, even if the quantum state of the two particles is entangled, it ought to be possible to specify a more complete description $\lambda$ of the joint state of the two particles, such that given $\lambda$, the probabilities can be written in the form
\begin{equation}\label{bipartitelocal}
P_{\lambda}(ab|XY) = P_{\lambda}(a|X) P_{\lambda}(b|Y).
\end{equation}
The state $\lambda$ is conventionally referred to as a \emph{hidden} state, since it is not part of the quantum description of the experiment. Any hidden state $\lambda$ which satisfies Eq.~(\ref{bipartitelocal}) is \emph{local}.
If the observed correlations $P(ab|XY)$ can be explained by a locally causal theory, then they can be written
\begin{equation}\label{bipartitelocalcorrelations}
P(ab|XY)=\sum_{\lambda}q_{\lambda}P_{\lambda}(a|X)P_{\lambda}(b|Y)
\end{equation}
with $q_\lambda\geq 0$ and $\sum_\lambda q_\lambda=1$. On the other hand, if correlations $P(ab|XY)$ violate a Bell inequality \cite{bellspeakable}, then they cannot be written in this form. Such correlations cannot be explained by a locally causal theory, and are referred to as nonlocal correlations.

Quantum nonlocality is a puzzling aspect of nature, but also an important resource for quantum information processing. An information theoretic interpretation of quantum nonlocality is that two separated parties who wish to simulate the experiment with classical resources cannot do so using only shared random data - they must also communicate with one another. The fact that entangled quantum states can produce nonlocal correlations enables the quantum advantage in communication complexity problems \cite{commcomplex}, device independent quantum cryptography \cite{bhk,diqkd}, randomness expansion \cite{dirng}, and measurement-based quantum computation \cite{oneway,oneway2}. 

With three or more systems, qualitatively different kinds of nonlocality can be distinguished. For definiteness, consider the tripartite case. If correlations can be written
\begin{equation}
P(abc|XYZ) = \sum_\lambda q_\lambda P_\lambda(a|X)\,P_\lambda(b|Y)\,P_\lambda(c|Z),
\end{equation}
with $0\leq q_\lambda\leq 1$ and $\sum_\lambda q_\lambda =1$, then they are local. Otherwise they are nonlocal. But, as pointed out by Svetlichny \cite{svet}, some correlations can be written in the hybrid local-nonlocal form
\begin{align}\label{svetlichnylocal}
&P(abc|XYZ) = \sum_\lambda q_{\lambda}\,P_\lambda(ab|XY)\,P_\lambda(c|Z) + \nonumber\\
& \sum_\mu q_{\mu}\,P_\mu(ac|XZ)\,P_\mu(b|Y) + \sum_\nu q_{\nu}\,P_\nu(bc|YZ)\,P_\nu(a|X),
\end{align}
where $0\leq q_{\lambda},q_{\mu},q_{\nu} \leq 1$ and $\sum_\lambda q_{\lambda}+\sum_\mu q_{\mu}+\sum_\nu q_{\nu}=1$. Here, each term in the decomposition factorizes into a product of a probability pertaining to one party's outcome alone, and a joint probability for the two other parties. We say that correlations of the form (\ref{svetlichnylocal}) are $S_2$-local. If correlations cannot be written in this form, then a term like $P_{\lambda}(abc|XYZ)$ must appear somewhere in the decomposition. Such correlations are often said to exhibit \emph{genuine 3-way} nonlocality, although we will refer to this as \emph{Svetlichny} nonlocality. Svetlichny introduced an inequality, violation of which implies Svetlichny nonlocality.
Svetlichny's inequality can be violated by appropriate measurements on a GHZ or W state \cite{mitchell}. 

In further work, Seevinck and Svetlichny \cite{seevinck}, and independently, Collins et al. \cite{collins}, generalized the tripartite notion of Svetlichny nonlocality to $n$ parties. 
In both Refs.~\cite{seevinck} and \cite{collins}, an inequality is derived that detects $n$-partite Svetlichny nonlocality. 
See also Refs.~\cite{mitchell,jones,bancal}.

The present work considers two alternative definitions of genuine multipartite nonlocality, which are different from Svetlichny's. We argue that these definitions are better motivated, both physically, and from the point of view of information theory. We show that the alternative definitions are strictly weaker than Svetlichny's and describe a Bell inequality such that its violation is sufficient for genuine $3$-way nonlocality according to both alternative definitions. Numerical evidence suggests that any pure, 3-way entangled quantum state can produce correlations that violate this inequality. On the other hand, there exist pure, 3-way entangled quantum states for which we have not been able to find any measurements giving rise to Svetlichny nonlocality.  

\paragraph{Different kinds of nonlocality.}
Consider again the case of bipartite correlations. There are various ways in which a hidden state $\lambda$ might fail to be local. Let $P_{\lambda}(a|XY) = \sum_b P_{\lambda}(ab|XY)$ be the marginal probability for Alice to obtain outcome $a$ when the measurement choices are $X$ and $Y$, and similarly let $P_{\lambda}(b|XY) = \sum_a P_{\lambda}(ab|XY)$ be the probability for Bob to obtain $b$. Suppose that $\lambda$ satisfies
\begin{align}
P_{\lambda}(a|XY) &= P_{\lambda}(a|XY') \quad \forall a,X,Y,Y' \label{nosigbobtoalice}\\
P_{\lambda}(b|XY) &= P_{\lambda}(b|X'Y) \quad \forall b,Y,X,X'.\label{nosigalicetobob}
\end{align} 
In this case, if Alice and Bob are in possession of two particles, which they know to be in the hidden state $\lambda$, then even if $\lambda$ is nonlocal, observing her own outcome gives Alice no information about Bob's measurement choice. This is because the marginal probabilities for $a$ are independent of Bob's choice. Hence Bob cannot send signals to Alice by varying his measurement choice. Similarly, Alice cannot send signals to Bob. Such a $\lambda$ is \emph{non-signalling}. 

If Eq.~(\ref{nosigbobtoalice}) is satisfied but Eq.~(\ref{nosigalicetobob}) is violated, then Bob's outcome gives him at least some information about Alice's measurement choice, hence Alice can send signals to Bob. The hidden state $\lambda$ is \emph{1-way signalling}. Similarly if Eq.~(\ref{nosigalicetobob}) is satisfied but Eq.~(\ref{nosigbobtoalice}) is violated. If Eqs.(\ref{nosigbobtoalice}) and (\ref{nosigalicetobob}) are both violated then $\lambda$ is \emph{2-way signalling}.  

\begin{figure}[tb]
\centering
\includegraphics[width=\linewidth]{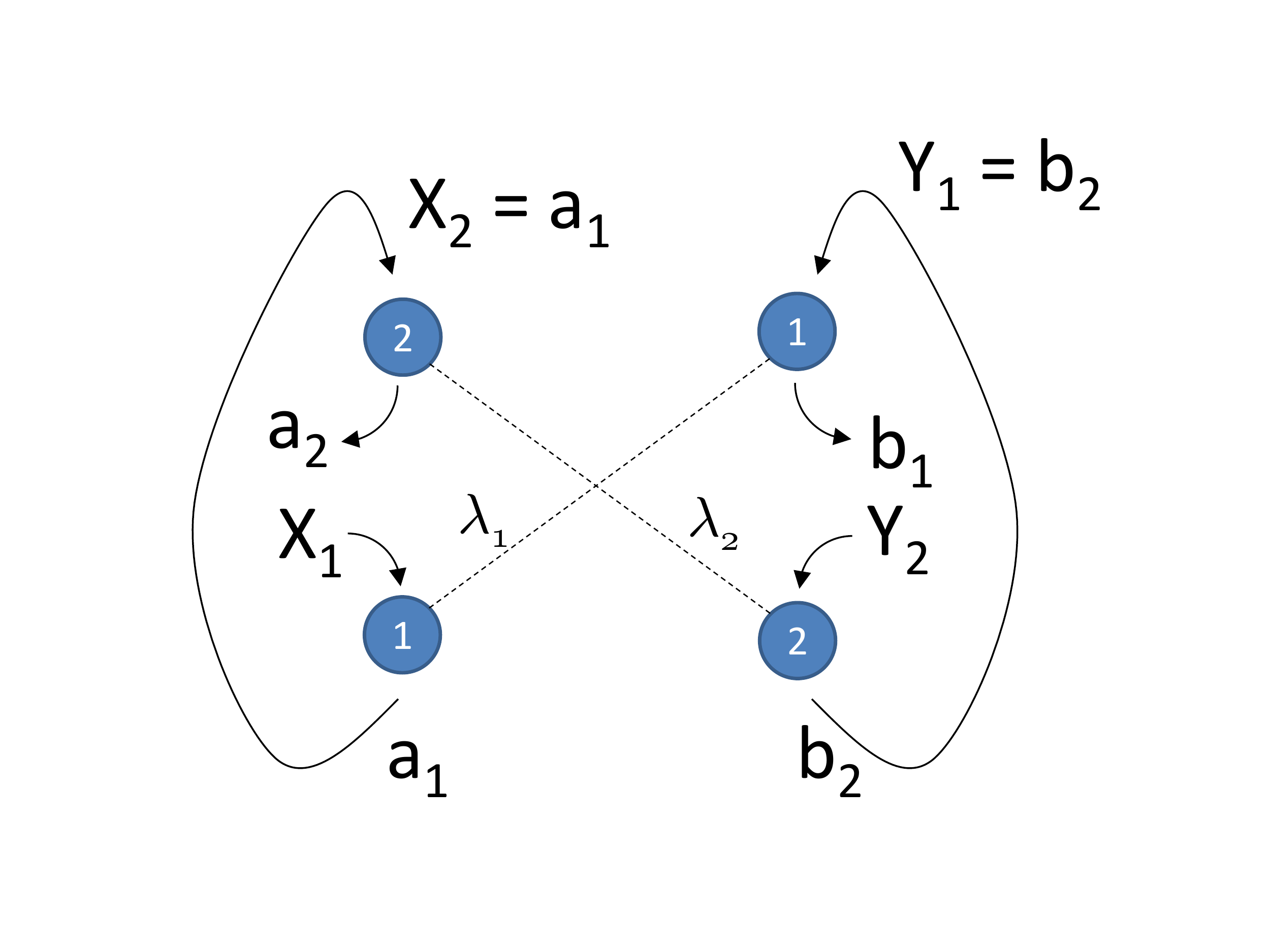}
\caption{Let $X,Y,a,b \in \{ 0,1 \}$. The particle pair labelled $1$ is independent from the pair labelled $2$. The joint state $\lambda_1$ is such that if $a_1=Y_1$, then $P_{\lambda_1}(a_1b_1|X_1Y_1)=0$, whereas $\lambda_2$ is such that if $b_2 \ne X_2$, then $P_{\lambda_2}(a_2b_2|X_2Y_2)=0$. Consistent predictions are impossible if measurement choices are as shown.}
\label{paradoxfigure2}
\end{figure}

So far, this discussion has followed many treatments of quantum nonlocality, in that no attention has been given to the timing of Alice's and Bob's measurements. It has been assumed -- naively -- that the measurements can unproblematically be regarded as simultaneous, or alternatively that the probabilities $P_{\lambda}(ab|XY)$ are independent of the timing of the measurements. With 1-way and 2-way signalling states, this can quickly cause problems. Suppose that a hidden state $\lambda$ is 1-way signalling from Alice to Bob. Then the outcome probabilities for a measurement of Bob's depend on which measurement setting Alice chooses. If Bob obtains his measurement outcome before Alice chooses her setting (with respect to some frame) then this implies some kind of backwards causality (with respect to that frame). Worse,  Figure~\ref{paradoxfigure2} shows how signalling hidden states can lead to grandfather-style paradoxes, where no consistent assignment of probabilities to outcomes is possible.

One solution to these problems would be to restrict attention to models that involve only non-signalling hidden states. But a more general solution is to introduce a notion of hidden state, according to which the outcome probabilities can vary according to the timing of the measurements.  In a fully general treatment, $\lambda$ will be time dependent, or alternatively, $\lambda$ will refer to the state of the particles at some fixed time (perhaps just after creation) in some fixed frame, and the probabilities for outcomes depend on the exact timing of the measurements. 

For now, let's keep things simple. Consider a hidden state $\lambda$ such that the probabilities do not depend on the exact timing of measurements, but do depend on the time \emph{ordering}, where this ordering is determined with respect to a fixed background frame. If Alice performs $X$ before Bob performs $Y$, the probabilities are given by
\begin{equation}\label{bipartitealicefirst}
P_{\lambda}^{A<B}(ab|XY).
\end{equation}  
If Bob performs $Y$ before Alice performs $X$, the correlations may be different with probabilities given by 
\begin{equation}\label{bipartitebobfirst}
P_{\lambda}^{B<A}(ab|XY).
\end{equation}
Paradoxes like that of Figure~\ref{paradoxfigure2} are avoided if:
\begin{enumerate}
\item the fixed background frame determining the time ordering of measurements is the same for all particle pairs,
\item the correlations $P_{\lambda}^{A<B}(ab|XY)$ and $P_{\lambda}^{B<A}(ab|XY)$ are at most 1-way signalling, with $P_{\lambda}^{A<B}(ab|XY)$ satisfying Eq.~(\ref{nosigbobtoalice}) and $P_{\lambda}^{B<A}(ab|XY)$ satisfying Eq.~(\ref{nosigalicetobob}).   
\end{enumerate}
An explicit model depending on the time ordering of measurements and satisfying the above two conditions is given by Bohm's theory \cite{bohm}.

Given a set of bipartite quantum correlations $P(ab|XY)$, these considerations about the time ordering of measurements do not make any difference to the basic question of whether $P(ab|XY)$ is nonlocal or not, which is perhaps why time ordering is not often emphasized. In the case of three or more observers, however, it makes an important difference to the classification of different kinds of multi-partite nonlocality.  

\paragraph{Genuine 3-way nonlocality.}
In Eq.~(\ref{svetlichnylocal}), the probabilities are assumed to be independent of the time ordering of measurements, and no constraint is placed on the bipartite correlations appearing in each term. So $P_{\lambda}(ab|XY)$, for example, can be 1-way or 2-way signalling. But as shown above, problems can arise with signalling hidden states, including paradoxes that can result if measurement outcomes can be used to determine measurement choices on other particles. One remedy is to consider only non-signalling hidden states. This suggests the following definition of genuine tripartite nonlocality.  
\begin{definition}\label{nonsiggennonloc}
Suppose that $P(abc|XYZ)$ can be written in the form 
\begin{align}\label{genuinenonsignonloc}
&P(abc|xyz) = \sum_\lambda q_\lambda P_{\lambda}(ab|xy)P_{\lambda}(c|z)+\nonumber \\ 
&\sum_\mu q_\mu P_{\mu}(ac|xz)P_{\mu}(b|y)+\sum_\nu q_\nu P_{\nu}(bc|yz)P_{\nu}(a|x),
\end{align}
where the bipartite terms are non-signalling, satisfying conditions of the form (\ref{nosigbobtoalice}) and (\ref{nosigalicetobob}). Then the correlations are $NS_2$-local. Otherwise, we say that they are genuinely 3-way NS nonlocal.
\end{definition}

As we have seen, however, a more general remedy is to define hidden states in such a way that correlations can depend on the time ordering of the measurements. It is convenient to write $P_{\lambda}^{T_{AB}}(ab|XY)$ for a set of time-order-dependent correlations, so that $P_{\lambda}^{T_{AB}}(ab|XY)=P_{\lambda}^{A<B}(ab|XY)$ when Alice measures before Bob and $P_{\lambda}^{T_{AB}}(ab|XY)=P_{\lambda}^{B<A}(ab|XY)$ when Bob measures before Alice. As always, assume that $P_{\lambda}^{A<B}(ab|XY)$ and $P_{\lambda}^{B<A}(ab|XY)$ are at most 1-way signalling, satisfying Eqs.~(\ref{nosigbobtoalice}) and (\ref{nosigalicetobob}) respectively. 
\begin{definition}\label{usgennonloc}
Suppose that $P(abc|XYZ)$ can be written in the form
\begin{align}\label{gnl}
&P(abc|xyz) = \sum_\lambda q_\lambda P_{\lambda}^{T_{AB}}(ab|xy)P_{\lambda}(c|z)+\\
&\sum_\mu q_\mu P_{\mu}^{T_{AC}}(ac|xz)P_{\lambda}(b|y)+\sum_\nu q_\nu P_{\nu}^{T_{BC}}(bc|yz)P_{\lambda}(a|x)\nonumber.
\end{align}
Then the correlations are $T_2$-local. Otherwise they are genuinely 3-way nonlocal.
\end{definition}

\paragraph{Interpretation from the point of view of quantum information.}
It is useful to contrast Definition 2 and Svetlichny's one from the perspective of classical simulations of quantum correlations in term of shared random data and communication (for examples of such a model, see Refs.~\cite{toner}). Svetlichny models naturally correspond to simulation models where all parties receive their input (the measurement they are to simulate) at the same time, then there are several rounds of communication between subsets of the parties, and finally, all parties produce an output (the measurement outcome). Models of the form (\ref{gnl}), on the other hand, correspond to simulation models where inputs are given to the parties in a sequence, where the order in the sequence  is arbitrary and not fixed in advance. On receiving an input, a party must produce an output immediately and may send a communication to a subset of the other parties. This means that although a party's output can depend on communications already received, it cannot depend on communications from parties later in the sequence. 

The distinction between both types of models is crucial for the simulation of quantum correlations in applications such as measurement-based computation where measurements are performed adaptively, that is, where the choice of which measurement to perform on a particular system may depend on the measurement outcome that was obtained from another system. In this context, simulation models \`a la Svetlichny in which all inputs are given at the same time are not relevant. 

Finally, models based on the definition (\ref{genuinenonsignonloc}) can be interpreted as simulation models where classical communication is replaced by no-signalling resources \cite{correlations} (such as PR boxes \cite{pr}). They are well adapted to the characterization of nonlocality for cryptographic applications secure against post-quantum adversaries \cite{bhk}. 

\paragraph{Characterization and detection of $3$-way nonlocality.}
Given the sets $NS_2$, $T_2$ and $S_2$, corresponding, respectively, to Definition 1, Definition 2, and Sveltichny definition, we have the following results (see details in Appendix). First, the different definitions of multipartite nonlocality are inequivalent, as one can show that $NS_2\subset T_2\subset S_2$ where the inclusions are strict (see Appendix C). {Note that contrarily to $S_2$ models, both $NS_2$ and $T_2$ models can only reproduce no-signalling correlations (this is true on average for $T_2$ models even though they may involve 1-way signalling between the parties at the hidden level, see Appendix B).}. Second, given correlations $P(abc|XYZ)$ with a finite number of measurement settings and outputs, it is a linear programming problem to determine whether they belong to the sets $NS_2$, $T_2$ or $S_2$ (see Appendix A). Furthermore, if correlations $P(abc|XYZ)$ are $NS_2$ or $T_2$ local, then (see Appendix B)
\begin{eqnarray} 
I &=& -2P(A_1B_1) - 2P(B_1C_1)-2P(A_1C_1) \nonumber \\
&& - P(A_0B_0C_1) - P(A_0B_1C_0) - P(A_1B_0C_0)\nonumber\\
&&+ 2P(A_1B_1C_0) + 2P(A_1B_0C_1) + 2P(A_0B_1C_1) \nonumber \\ 
&& + 2P(A_1B_1C_1) \leq 0,  \label{ineq}
\end{eqnarray}
where we have introduced the notation
$P(A_i B_j)\equiv P(a=0, b=0|X=i, Y=j)$, $P(A_i B_j C_k) \equiv P(a=0, b=0, c=0|X=i, Y=j, Z=k)$.
Just as Svetlichny introduced an inequality, violation of which implies Svetlichny nonlocality, Eq.(\ref{ineq}) is a Bell-type inequality, violation of which implies that correlations are 3-way (NS and T) nonlocal. {In Appendix D we provide also a complete characterization of the $NS_2$ polytope in presence of binary inputs and outputs. Inequality \eqref{ineq} belongs to the family number 6 in this list, and is thus a tight constraint on the $NS_2$ as well as the $T_2$ sets.}

\paragraph{Multipartite nonlocality and noisy quantum states.}
It is interesting to investigate the extent to which different quantum states can produce each type of multipartite nonlocality. Consider an experiment in which measurements are performed on a tripartite quantum state, with each party having a choice of two measurement settings, and each measurement having two possible outcomes. 
Let 
\begin{align}
|GHZ\rangle &= 1/\sqrt{2}(|000\rangle + |111\rangle), \\
|W\rangle &= 1/\sqrt{3}(|001\rangle + |010\rangle + |100\rangle), \\ 
\rho_{GHZ} &= p\, |GHZ\rangle\langle GHZ| + (1-p) I/8, \\
\rho_W &= p\, |W\rangle\langle W| + (1-p) I/8,
\end{align}
where $I$ is the identity and $0\leq p \leq 1$. We have determined using linear programming the minimum values of $p$ for which the states $\rho_{GHZ}$ and $\rho_W$ will exhibit each kind of multipartite nonlocality. Results are summarized in Table~\ref{thresholdtable}.
\begin{table}
\begin{tabular}{| c | c | c | c |}
\hline
State & $p_{NS}$ & $p_{T}$ & $p_{S}$ \\
\hline
$\rho_{GHZ}$ & $1/\sqrt{2}$ & $1/\sqrt{2}$ & $1/\sqrt{2}$ \\
$\rho_W$ & $0.8$ & $0.82$ & $0.92$ \\ 
\hline
\end{tabular}\caption{The table shows the minimum values of the $p$ parameter required for the quantum states $\rho_{GHZ}$ and $\rho_W$ to exhibit genuine multipartite nonlocality. {These values were found by numerical optimization}. It is assumed that three parties each have two possible measurement settings, each with two outcomes. If $p>p_{NS}$ then correlations can be produced which are 3-way NS-nonlocal (see Definition~\ref{nonsiggennonloc}). If $p> p_{T}$, then correlations can be produced which are 3-way nonlocal (Definition~\ref{usgennonloc}). If $p>p_{S}$, then correlations can be produced which are Svetlichny nonlocal. { Inequalities demonstrating the different notions of nonlocality of $\rho_W$ for values of $p$ higher that the above thresholds are described in Appendix D.}}\label{thresholdtable}
\end{table}
For the noisy GHZ state, it makes no difference which definition is employed -- 3-way NS-nonlocal, 3-way nonlocal, and Svetlichny nonlocal correlations can be generated whenever $p>1/\sqrt{2}$. In the case of the noisy $W$ state, there is a range of values of $p$ for which the state is too noisy to exhibit Svetlichny nonlocality, but can still produce correlations which are 3-way nonlocal, and similarly a range of values of $p$ for which the state is too noisy to exhibit 3-way nonlocality, but can still produce correlations which are 3-way NS-nonlocal. This again demonstrates that the different definitions of multipartite nonlocality are strictly inequivalent.

\paragraph{Multipartite nonlocality and tripartite entanglement.}
Finally, we conclude by presenting numerical results that suggests that all pure tripartite entangled states are three-way nonlocal. An arbitrary pure state of three qubits that is genuinely tripartite entangled can always be written in the form \cite{trient} $|\psi\rangle=\lambda_0|000\rangle+\lambda_1e^{\phi}|100\rangle+\lambda_2|101\rangle+\lambda_3|110\rangle+\lambda_4|111\rangle$, with $\phi\in[0,\pi]$, $\lambda_i\geq 0$, $\sum_{i}\lambda_i^2=1$, $\lambda_0\neq 0$, $\lambda_2+\lambda_4\neq 0$ and $\lambda_3+\lambda_4\neq 0$. We tested inequality (\ref{ineq}) for $8^5=32768$ states of this form obtained by considering 8 possible values for 5 independent variables parametrizing these states. After numerical optimization of the measurement settings, a violation was found in each case. We thus conjecture that all pure tripartite entangled states are three-way nonlocal.
Note, however, that we were not able to find any violation of the Svetlichny type for the following tripartite entangled state $|\psi\rangle=\frac{\sqrt{3}}{2}|000\rangle+\frac{\sqrt{3}}{4}|110\rangle+\frac{1}{4}|111\rangle$ (though it violates inequality (\ref{ineq})).  Our search included the {1087} different Svetlichny inequalities introduced in \cite{symbell}, as well as a linear programming search over the Svetlichny polytope with two measurements settings per party.

\paragraph{Note added.} While the present manuscript formally makes public the definitions and results presented here, they have already been communicated privately to close collaborators. In particular, Definitions 1 and 2 where used in the following Refs \cite{almeida, pironio, bancal2}. Note also the independent work \cite{chicos} where Definition 2 is introduced and motivated from a different (though related) perspective.

\paragraph{Acknowledgments.} This work was supported by the EU FP7 QCS project, the CHIST-ERA DIQIP project, the Swiss NCCRs QSIT, the Interuniversity Attraction Poles Photonics@be Programme (Belgian Science Policy), the Brussels-Capital Region through a BB2B Grant, {the National Research
Foundation and the Ministry of Education of Singapore}, the FQXi large grant "Time and the Structure of Quantum Theory", and the FRFC DIQIP project.

\section{Appendix A}
Here we show that the sets $NS_2$, $T_2$ and $S_2$ can be characterized by linear constraints. This implies that linear programming can be used to decide if a set of correlations $P(abc|XYZ)$ belongs to the sets $NS_2$, $T_2$ and $S_2$ or to compute the maximal value that a Bell expression $\sum_{abcXYZ} C_{abcXYZ} P(abc|XYZ)$ for each of these sets.

We start with $S_2$-local correlations defined in Eq.~\eqref{svetlichnylocal}. In this decomposition, the bipartite correlations $P_\lambda(ab|XY)$ are arbitrary conditional probability distributions. Such distributions can always be expressed as a convex combination of deterministic strategies of the form $P^*(ab|XY) = \delta(a,a^*(X,Y))\delta(b,b^*(X,Y))\in\{0,1\}$ where $a^*,b^*$ can take any value admissible for $a$ and $b$. Namely, one can always write $P_\lambda (ab|XY)=\sum_i q_{i|\lambda} P_i^*(ab|XY)$ with some positive normalized weights $q_{i|\lambda}$ satisfying $q_{i|\lambda}\geq 0$ and $\sum_i q_{i|\lambda}=1$. Such a decomposition exists in general for arbitrary $n$-partie correlations. For instance, single-party correlations can also be decomposed as $P_\lambda(c|Z)=\sum_{j|\lambda} q_{j|\lambda} P_j^*(c|Z)$ with positive normalized weights $q_{j|\lambda}$ and $P_{j|\lambda}^*(c|Z)\in\{0,1\}$ deterministic. The first term in Eq.~\eqref{svetlichnylocal} of the main text can thus be written as 
\begin{equation}\label{decompS2}
\begin{split}
\sum_\lambda q_\lambda P_\lambda(ab|XY)P_\lambda(c|Z)&=\sum_{\lambda,i,j} q_\lambda q_{i|\lambda} q_{j|\lambda} P_{i}^*(ab|XY) P_{j}^*(c|Z)\\
&=\sum_{i,j} q_{ij} P_{i}^*(ab|XY) P_{j}^*(c|Z)\\
&=\sum_{\lambda'}q_{\lambda'}P_{\lambda'}^*(ab|XY)P_{\lambda'}^*(c|Z),
\end{split}
\end{equation}
where we set $\lambda'=(i,j)$ and $q_{\lambda'}=\sum_\lambda q_\lambda q_{i|\lambda}q_{j|\lambda}$. Since all $q_\lambda, q_{i|\lambda}, q_{j|\lambda}$ are normalized and $i$ and $j$ are independent given $\lambda$, $q_{\lambda'}$ satisfies $q_{\lambda'}\geq 0$, $\sum_{\lambda'}q_{\lambda'}=1$. A similar decomposition holds for the two other terms of Eq.~\eqref{svetlichnylocal}. It is thus sufficient to consider only product of deterministic strategies in decomposition~\eqref{svetlichnylocal}.

Since the number of deterministic strategies is finite when all parties have a finite number of possible inputs and outputs, we see that the correlations $P(abc|XYZ)$ admits a $S_2$ model if they can be written as
\begin{equation}
\begin{split}
P(abc|XYZ)=&\sum_\lambda q_\lambda P_\lambda^*(ab|XY)P_\lambda^*(c|Z)\\
&+ \sum_\mu q_\mu P_\mu^*(ac|XZ)P_\mu^*(b|Y)\\
& + \sum_\nu q_\nu P_\nu^*(bc|YZ)P_\nu^*(a|X)\\
\text{with\ \,} q_\lambda,q_\mu,q_\nu\geq&\,0\,, \ \ \sum_\lambda q_\lambda + \sum_\mu q_\mu + \sum_\nu q_\nu = 1,
\end{split}
\end{equation}
where $P^*$ are deterministic strategies. This represents a set of linear constraints on the probability weights $q_\lambda$, $q_\mu$, $q_\nu$.

In contrast with the $S_2$ case, bipartite correlations $P_\lambda(ab|XY)=P_\lambda^{T_{AB}}(ab|XY)$ in the definition of $T_2$-local correlations can be at most 1-way signalling. Depending on the order in which measurements are performed they must thus possess a decomposition either as $P_\lambda^{A<B}(ab|XY)$ or as $P_\lambda^{B<A}(ab|XY)$. Using equations \eqref{nosigbobtoalice} and \eqref{nosigalicetobob} and a similar trick as before, these bipartite correlations can also be decomposed in terms of deterministic strategies either as $P_\lambda^{A<B}(ab|XY)=\sum_\sigma q_{\sigma|\lambda} P_\sigma^*(a|X)P_\sigma^*(b|XY)$ or $P_\lambda^{B<A}(ab|XY)=\sum_{\sigma'} q_{\sigma'|\lambda} P_{\sigma'}^*(a|XY)P_{\sigma'}^*(b|Y)$ where $0\leq q_{\sigma|\lambda},q_{\sigma'|\lambda}\leq 1$, $\sum_\sigma q_{\sigma|\lambda}=\sum_{\sigma'} q_{\sigma'|\lambda}=1$.

That a decomposition \eqref{gnl} exists for the time orderings $T_1=A<B<C$ and $T_2=B<A<C$ implies that the following condition holds:
\begin{equation}\label{decompPab}
\sum_\lambda q_\lambda P_\lambda^{A<B}(ab|XY)P_\lambda(c|Z) = \sum_\lambda q_\lambda P_\lambda^{B<A}(ab|XY)P_\lambda(c|Z).
\end{equation}
We refer to this quantity as $P^{AB/C}(abc|XYZ)$, and similarly define the quantities $P^{AC/B}(abc|XYZ)$, $P^{BC/A}(abc|XYZ)$. We have $P(abc|XYZ)=P^{AB/C}(abc|XYZ)+P^{AC/B}(abc|XYZ)+P^{BC/A}(abc|XYZ)$. 

Inserting the above decompositions of $P_\lambda^{A<B}(ab|XY)$ and $P_\lambda^{B<A}(ab|XY)$ in term of deterministic strategies in (\ref{decompPab}) , we find
\begin{equation}
\begin{split}
P^{AB/C}(abc|XYZ)=&\sum_{\lambda,\sigma} q_\lambda q_{\sigma|\lambda} P_\sigma^*(a|X)P_\sigma^*(b|XY) P_\lambda^*(c|Z)\\
&=\sum_{\lambda,\sigma} q_{\lambda,\sigma} P_\sigma^*(a|X)P_\sigma^*(b|XY) P_\lambda^*(c|Z)
\end{split}
\end{equation}
and
\begin{equation}
\begin{split}
P^{AB/C}(abc|XYZ)=&\sum_{\lambda,\sigma'} q_\lambda q'_{\sigma'|\lambda} P_{\sigma'}^*(a|XY)P_{\sigma'}^*(b|Y) P_\lambda^*(c|Z)\\
&=\sum_{\lambda,\sigma'} q'_{\lambda,\sigma'} P_{\sigma'}^*(a|XY)P_{\sigma'}^*(b|Y) P_\lambda^*(c|Z)
\end{split}
\end{equation}
where $q_{\lambda,\sigma}$ and $q'_{\lambda,\sigma'}$ are arbitrary but satisfy $q_{\lambda,\sigma}, q'_{\lambda,\sigma'}\geq 0$, $\sum_\sigma q_{\lambda,\sigma}=\sum_{\sigma'} q'_{\lambda,\sigma'}=q_\lambda$. 

Altogether, we can write
\begin{widetext}
\begin{eqnarray}
\label{t2begin}
&&P(abc|XYZ) = P^{AB/C}(abc|XYZ) + P^{AC/B}(abc|XYZ) + P^{BC/A}(abc|XYZ)\\
&&P^{AB/C}(abc|XYZ)=\sum_{\lambda,\sigma} q_{\lambda,\sigma} P_{\sigma}^*(a|X)P_{\sigma}^*(b|XY) P_{\lambda}^*(c|Z)=\sum_{\lambda,\sigma'} q'_{\lambda,\sigma'} P_{\sigma'}^*(a|XY)P_{\sigma'}^*(b|Y) P_{\lambda}^*(c|Z)\label{ABconstr}\\
&&P^{AC/B}(abc|XYZ)=\sum_{\mu,\tau} q_{\mu,\tau} P_{\tau}^*(a|X)P_{\tau}^*(b|XY) P_{\mu}^*(c|Z)=\sum_{\mu,\tau'} q'_{\mu,\tau'} P_{\tau'}^*(a|XY)P_{\tau'}^*(b|Y) P_{\mu}^*(c|Z)\label{ACconstr}\\
&&P^{BC/A}(abc|XYZ)=\sum_{\nu,\kappa} q_{\nu,\kappa} P_{\kappa}^*(a|X)P_{\kappa}^*(b|XY) P_{\nu}^*(c|Z)=\sum_{\nu,\kappa'} q'_{\nu,\kappa'} P_{\kappa'}^*(a|XY)P_{\kappa'}^*(b|Y) P_{\nu}^*(c|Z)\label{BCconstr}\\
&& q_{\lambda,\sigma},q'_{\lambda,\sigma'},q_{\mu,\tau},q'_{\mu',\tau'},q_{\nu,\kappa},q'_{\nu,\kappa'} \geq 0\\
&& \sum_\sigma q_{\lambda,\sigma}=\sum_{\sigma'} q'_{\lambda,\sigma'}=q_\lambda\,,\ \ \sum_\tau q_{\mu,\tau} = \sum_{\tau'} q'_{\mu,\tau'} =q_{\mu}\,,\ \ \sum_\kappa q_{\nu,\kappa} = \sum_{\kappa'} q'_{\nu,\kappa'}= q_{\nu},\\
&&\sum_\lambda q_\lambda+\sum_\mu q_\mu +\sum_\nu q_\nu=1\label{t2end}
\end{eqnarray}
\end{widetext}
which represents a set of linear constraints on the probability weights $q_{\lambda,\sigma}$, $q'_{\lambda,\sigma'}$, $q_{\mu,\tau}$, $q'_{\mu,\tau'}$, $q_{\nu,\kappa}$, $q'_{\nu,\kappa'}$.

Note that in the $T_2$ decomposition the correlations $P_\lambda(ab|XY)$ might not be identical if $A<B$ or $B<A$; only condition~\eqref{decompPab} is required to hold. In fact, this possibility distinguishes $T_2$-local correlations from $NS_2$-local ones for which $P_\lambda(ab|XY)$ is defined independently of the measurement order. $NS_2$-local correlations can thus be characterized by adding the conditions that 
\begin{equation}
\sum_\sigma q_{\lambda,\sigma} P_{\sigma}^*(a|X)P_{\sigma}^*(b|XY)=\sum_{\sigma'} q'_{\lambda,\sigma'} P_{\sigma'}^*(a|XY)P_{\sigma'}^*(b|Y)
\end{equation}
and similar ones for the $P^{AC/C}$ and $P^{BC/A}$ terms.

On the other hand, if we relax some of the constraints above, removing the right constraints of Eqs.~\eqref{ABconstr}-\eqref{BCconstr} for instance, one can obtain a linear program tuned to describe correlations that can be achieved when the order in which parties are measured is known ($A<B<C$ in this case). Indeed if it is known prior to measurement that particles will be measured in the order $A<B<C$, a model could use this information to propare correlations achievable in this order but not in different ones. The additional constraints are then not required to hold. More generally, correlations that can be produced when the measurement order is known beforehand, and for all possible orders, need only satisfy 
\begin{equation}
\begin{split}
P^{AB/C}(abc|XYZ)&=\sum_{\lambda,\sigma} q_\lambda P_{\lambda}^*(a|X)P_{\lambda}^*(b|XY) P_{\lambda}^*(c|Z)\\
&=\sum_{\lambda'} q_{\lambda'} P_{\lambda'}^*(a|XY)P_{\lambda'}^*(b|Y) P_{\lambda'}^*(c|Z)
\end{split}
\end{equation}
instead of Eqs.~\eqref{ABconstr}-\eqref{BCconstr} above (and similar conditions for the $P^{AC/B}$ and $P^{BC/A}$ terms). We refer to such models are $K_2$-local (see also~\cite{extremalBoxes}). Clearly, one has that $T_2\subseteq K_2 \subseteq S_2$. Using the presented programs one can show that these inclusions are strict.

\section{Appendix B}

Here we provide a proof that inequality (11) is satisfied by $T_2$ correlations.

\emph{Proof of inequality (11).}
First, we prove that all $T_2$ correlations are no-signalling. This justifies the use of marginal probabilities in equation (11).

If correlations $P_{\lambda}(abc|XYZ)$ are $T_2$ local, then they can be written in the form (\ref{gnl}). Let $P^{AB/C} \equiv \sum_{\lambda} q_{\lambda} P^{T_{AB}}_{\lambda}(ab|XY)P_{\lambda}(c|Z)$. Similarly, let $P^{AC/B}$ denote the second term on the right hand side of Eq.(\ref{gnl}) and $P^{BC/A}$ the third one. With the ordering $A<B<C$, the term $P^{AB|C}$ can be signalling only from Alice to Bob, the term $P^{AC|B}$ from Alice to Charlie and $P^{BC|A}$ from Bob to Charlie. In the case the ordering is $B<A<C$, the first term can be signalling this time only from Bob to Alice, but the two other terms must be identical. Since the left-hand-side of Eq.~\eqref{gnl} is independent of time ordering, the term $P^{AB/C}$ must be the same in both cases as well, which implies that it is non-signalling. By a similar argument one can show that the two other terms are also no-signalling, and so is any $T_2$ correlations.

Now, to show the validity of (\ref{ineq}), it is sufficient to show that it is satisfied by each term $P^{AB/C}$, $P^{AC/B}$, $P^{BC/A}$ separately. But the inequality is symmetric under permutations of the parties, hence it is sufficient to show that it is satisfied by $P^{AB/C}$. 

Consider the expression
\begin{eqnarray}\label{1wayineq} 
I_{A<B} &=& - P(A_1B_1) - P(B_1C_1|A_0) - P(A_1C_1) \nonumber \\
&-& \frac{1}{2}P(A_0B_0C_1) - P(A_1B_0C_0) + P(A_1B_1C_0) \nonumber\\
&+& P(A_1B_0C_1) + P(A_0B_1C_1)+ P(A_1B_1C_1),
\end{eqnarray}
where $P(B_jC_k|A_i) \equiv P(b=0,c=0|X=i,Y=j,Z=k)$.
Similarly, let $I_{B<A}$ be defined by an expression similar to that of Eq.~(\ref{1wayineq}), but with $A$ and $B$ interchanged.
For no-signalling distributions, one can check that $I=I_{A<B}+I_{B<A}$. The value of $I$ achievable with $P^{AB/C}$ can thus be computed as
\begin{equation}
\begin{split}
I(P^{AB/C}) &= I_{A<B}(P^{AB/C}) + I_{B<A}(P^{AB/C})\\
 &= I_{A<B}(P^{A<B/C}) + I_{B<A}(P^{B<A/C})
\end{split}
\end{equation}
where 
\begin{equation}\label{abeforebform}
P^{A<B/C}(abc|XYZ)=\sum_{\lambda} q_{\lambda} P_{\lambda}(a|X) P_{\lambda}(b|XY) P_{\lambda}(c|Z).
\end{equation}
and $P^{B<A/C}$ is similarly defined. 
It is easily verified that for distributions of the form~\eqref{abeforebform}, the bound $I_{A<B}\leq 0$ holds (for example by considering all deterministic strategies involving signalling from Alice to Bob). By symmetry, we also have that $I_{B<A}\leq 0$ for all $P^{B<A/C}$, and thus $I(P^{AB/C})\leq 0$, which concludes the proof.\hfill$\Box$

One can check that the $NS_2$ bound of inequality (11) coincides with the $T_2$ one, even though $NS_2$ correlations are in general less powerful. This inequality thus doesn't distinguish these two definitions. In contrast, we present in Appendix D several inequalities which have a different bound for each definition introduced in the main text (see Table III).

\section{Appendix C}

Here we show that the following strict inclusion holds: $NS_2\subset T_2\subset S_2$. Note that from the definitions of these sets provided in the main text, it is clear that $NS_2\subseteq T_2\subseteq S_2$: decomposition (9) is a special case of (10), which is a special case of (4). We thus only need to show here that these inclusions are strict. After that we also give an explicit example of quantum correlations which are genuinely 3-way nonlocal but do not violate any Svetlichny inequality.

\emph{Proof that definitions $NS_2$ and $T_2$ are inequivalent.} It is sufficient to describe a set of correlations that can be written in the form (\ref{gnl}) but not in the form (\ref{genuinenonsignonloc}). The following provides an example of such correlations. Suppose that (i) measurement choices and outcomes are binary-valued, (ii) marginal probabilities for a single party are always 1/2, (iii) the outcomes satisfy the following relations:
\begin{equation}\label{corr1}
\begin{split}
a_0+b_1=0,\ \ &\ \ a_1+c_0 = 0 \\
b_0+c_1=0,\ \ &\ \ a_0+b_0+c_0 = 0 \\
a_1+b_1+c_1=1
\end{split}
\end{equation}
where $a_x$ is Alice's outcome when her measurement choice is $x$, and similarly for the other parties. All sums are modulo 2.  We now show that these correlations admit a decomposition of the form of (\ref{gnl}), but not a decomposition of the form of (\ref{genuinenonsignonloc}).

Note first of all that these correlations are nonlocal: there is no possible assigment of definite values to $a_X,b_Y,c_Z$ such that each is either $0$ or $1$ and all the equations are satisfied. This can be seen by summing the right-hand sides and left-hand sides of the equations.

Let \emph{PR} correlations \cite{pr,kt} be non-signalling bipartite correlations with $a,b,X,Y$ binary, such that marginal probabilities are equal to $1/2$, and, up to relabelling of outcomes and measurement choices, the outcomes satisfy
\begin{equation}\label{pr}
a\oplus b = XY.
\end{equation}
It is known \cite{correlations} that binary non-signalling correlations $P(ab|XY)$ can always be written as a mixture of local terms and PR correlations. It follows that if some correlations are $NS_2$-local, then they can be written as
\begin{align}\label{qnsdecomp}
\sum_{\lambda} q_{\lambda} &P_{\lambda}(a|X) P_{\lambda}(b|Y) P_{\lambda}(c|Z) + \nonumber \\
 & \sum_{\mu} q_{\mu} P^{PR}_{\mu}(ab|XY) P_{\mu}(c|Z) + \mathrm{perms.}, 
\end{align}
where $P^{PR}_{\mu}(ab|XY)$ are $PR$ correlations. Consider a term with PR correlations shared between Alice and Bob. Alice's marginal probabilities are $1/2$ and her outcome uncorrelated with Charles. But this contradicts Eq.~(\ref{corr1}). Similar arguments with parties permuted imply that the weight of the nonlocal terms in Eq.~(\ref{qnsdecomp}) must be zero. But we saw that the correlations \eqref{corr1} cannot be written as a sum of only local terms only since they are nonlocal. The are thus not $NS_2$-local.

The correlations \eqref{corr1} are, however, $T_2$-local according to Definition~\ref{usgennonloc}, since they can be written in the form
\[P_Q(abc|xyz)=\sum_\lambda q_{\lambda} P_\lambda(a|x)P_{\lambda}^{T_{BC}}(bc|xy).\]
Here is an explicit decomposition of this form for these correlations. Let $\lambda$ be a pair of variables $(\lambda_0,\lambda_1)$ with $\lambda_0,\lambda_1\in\{0,1\}$ and $q_{\lambda_0,\lambda_1} = 1/4$. Alice's outcomes are given by
\begin{equation}
a_0 = \lambda_0,\quad a_1=\lambda_1.
\end{equation}
If Bob measures before Charles,
\begin{eqnarray}
b_0=\lambda_0+\lambda_1, &\,& b_1=\lambda_0\\
c_0=\lambda_1,&\,& c_1=\lambda_0+\lambda_1+Y
\end{eqnarray}
If Charles measures before Bob,
\begin{eqnarray}
b_0=\lambda_0+\lambda_1+Z, &\,& b_1=\lambda_0\\
c_0=\lambda_1,&\,& c_1=\lambda_0+\lambda_1+1.
\end{eqnarray}
One can check that these correlations fulfill the conditions \eqref{corr1} for both time orderings $B<C$ and $C<B$. They are thus $T_2$-local.\hfill$\Box$

\emph{Proof that $T_2$ and $S_2$ are inequivalent.}
Since we already proved the validity of inequality (11) for $T_2$-local correlations in appenix B, it is sufficient to show that this inequality can be violated by non-signalling $S_2$ correlations. For this, let us consider the following four strategies:
\begin{align*}
\text{1)\ \ }&a_X=X+Z-XZ,\ \ b_Y=0,\ \ c_Z=1,\\
\text{2)\ \ }&a_X=1-Z+XZ,\ \ b_Y=Y,\ \ c_Z=1,\\
\text{3)\ \ }&a_X=0,\ \ b_Y=Y-YZ,\ \ c_Z=1-Z,\\
\text{4)\ \ }&a_X=1-X,\ \ b_Y=1-Y+YZ,\ \ c_Z=Z.
\end{align*}
In strategies 1) and 2) Alice's outcome depends on Charlie's input, but otherwise only local inputs are required. Similarly, in the last two strategies, only Bob requires knowledge of Charles' inputs. These are thus valid $S_2$-local strategies.

While each of these strategies is signalling, one can check that their uniform mixture satisfies
\begin{equation}
\begin{split}
&\langle a_X\rangle = 1/2,\ \ \langle b_Y\rangle = (1+Y)/4,\ \ \langle c_Z\rangle = 3/4\\
&\langle a_X b_Y\rangle = (1-X+XY)/4\\
&\langle a_Xc_Z\rangle = (1+X+Z-XZ)/4\\
&\langle b_Yc_Z\rangle = (2Y+Z-YZ)/4\\
&\langle a_Xb_Yc_Z\rangle = (Y+Z-XZ-YZ+XYZ)/4\\
\end{split}
\end{equation}
and is thus no-signalling. Evaluation of expression (11) on these correlations yields the value $1/4>0$, thus violating the inequality.\hfill$\Box$
\ \\

Let us now give an example of quantum correlations which are 3-way nonlocal, but do not violate any Svetlichny inequality. For this, we define the following inequality:
\be\label{oldineq}
\av{A_0B_0} + \av{A_0C_0}+ \av{B_0C_1}-\av{A_1B_1C_0}+\av{A_1B_1C_1} \leq 3.
\ee
One can check that the inequality is valid for $T_2$-local correlations (with the linear programs of appendix A for instance). Here $\av{A_XB_Y}=\sum_{ab} (-1)^{a+b}P(ab|XY)$, $\av{A_XB_YC_Z}=\sum_{abc} (-1)^{a+b+c}P(abc|XYZ)$.

Now consider the correlations obtained with the following measurements on a GHZ state $|GHZ\rangle = (|000\rangle+|111\rangle)/\sqrt{2}$. Alice: $0\rightarrow \sigma_z, 1\rightarrow \sigma_x$. Bob: $0\rightarrow \sigma_z, 1\rightarrow \sigma_x$. Charles: $0\rightarrow 1/\sqrt{2}(\sigma_z+\sigma_x), 1\rightarrow 1/\sqrt{2}(\sigma_z-\sigma_x)$. With these measurements, the left hand side of Eq.~(\ref{oldineq}) is equal to $1+2\sqrt{2}$, thus the inequality is violated. Yet, these correlations do not violate the Svetlichny inequality, or indeed any Svetlichny-type inequality, as they are Svetlichny local. This is shown below by writing down an explicit model of the form of Eq.~(\ref{svetlichnylocal}) (with some terms signalling).

In our model, Alice is always local and Charles communicates to Bob. In each run, Charles outputs either $0$ or $1$ with probability $1/2$ (whatever be his input) according to a strategy fixed in advance and known to all parties. If Charles output $c=0$, Alice and Bob end up in the state $\cos(\pi/8) |00\rangle+e^{iz\pi}\sin(\pi/8)|11\rangle$ where $z$ is the value of Charles's inputs. Now, the correlations observed by Alice and Bob when they perform measurements $\sigma_z,\sigma_x$ on this state are local and there exists a local model where Alice's strategy is the same for $z=0,1$. This model uses two shared random bits $r_0,r_1$ where $r_0$ is distributed according to $P(r_0=0)=\cos^2(\pi/8)$,  $P(r_0=1)=\sin^2(\pi/8)$, and $r_1$ is uniform. Alice outputs $r_x$ and Bob $r_0+y(r_1+z)$. Thus we see that only Bob needs to receive Charles's input. Similarly if Charles output $c=1$, Alice and Bob end up in the state $\sin(\pi/8) |00\rangle+e^{iz\pi}\cos(\pi/8)|11\rangle$ and Alice and Bob can reproduce the corresponding correlations if they output $r_x+x+1$ and $r_0+1+y(r_1+z)$.

\section{Appendix D}\label{listOfIneqs}

Here we provide a complete characterization of the tripartite $NS_2$ polytope for two binary inputs per party. We also mention which inequalities can be used to recover the violations reported in table I of the main text.

As mention in appendix C, the extremal points of the $NS_2$ set, in presence of binary inputs and outcomes, are combinations of deterministic strategies with PR correlations. They can thus be listed easily. Using the polytope software porta~\cite{porta} we found all 405056 facets of the $NS_2$ polytope from this list of extremal points. Under permutation of parties, inputs and outputs, these facets get distributed within 185 different families which are listed in table~\ref{tvertphys}. Here we use the notation $\langle A_X\rangle = \sum_a (-1)^a P(a|X)$, $\langle A_X B_Y\rangle = \sum_{ab}(-1)^{a+b} P(ab|XY)$, etc.

The first family listed in table~\ref{tvertphys} represents the positivity of probabilities, i.e. $P(abc|XYZ) \geq 0$, whereas the last one is Svetlichny's original inequality~\cite{svet}, and inequality~\eqref{ineq} of the main text belongs to class number 6. Note that since the Svetlichny inequality is a facet of the $NS_2$-local polytope, and it is of course satisfied by $S_2$-local correlations, it must be a facet of the $T_2$ and $S_2$ sets of correlations as well.

For all inequalities listed here we computed also the $T_2$ and $S_2$ bounds with the aid of the programs described in Appendix A. These values are reported in Table~\ref{tbounds}, together with the maximum quantum violations. These violations were found by considering measurements on states of three qubits. They were checked to be optimal up to $\sim10^{-7}$ using the NPA hierarchy~\cite{NPA} up to local level 5, except for inequality 137 for which a small gap remains.

Considering measurements on the three-qubit W state $|W\rangle=1/\sqrt{3}(|001\rangle+|010\rangle+|100\rangle)$, one can check that inequalities in the class number 138 can be violated up to 12.4862. This certifies that states of the form $\rho_W = p\, |W\rangle\langle W| + (1-p) I/8$ can be non-$NS_2$-local as soon as $p\geq 10/12.4862= 0.8009$, in agreement with Table~\ref{thresholdtable} of the main text. Similarly, one can check that inequality number 12 in Table~\ref{tvertphys} can be violated up to 7.3137 by measuring a W state. Since the $T_2$ bound of this inequality coincides with the $NS_2$ bound (c.f. Table~\ref{tbounds}), this implies that states of the form $\rho_W$ are non-$T_2$-local for $p\geq0.8204$, in agreement with Table~\ref{thresholdtable}. Finally, the bound for $S_2$-locality of the W state can be found by checking that inequality in the class 185 can be violated up to 4.3546 by the W state.

\setlength{\extrarowheight}{0.4em}

\begin{longtable*}{|c|p{15cm}|}
\caption{Complete characterization of the tripartite $NS_2$ polytope with binary inputs and outcomes. One representative facet is described here for every family of inequalities which are equivalent to each other under relabelling of parties, inputs and outcomes.}\label{tvertphys}\\
\hline
Class & \hspace{6cm} Representative\\
\hline
\endfirsthead
\caption{continued}\\
\hline
Class & \hspace{6cm} Representative\\
\hline
\endhead
  \multicolumn{2}{l}{{Continued on the next page...}}
\endfoot
  \hline
\endlastfoot

1 & $ - \langle A_1\rangle  - \langle B_1\rangle  - \langle A_1B_1\rangle  - \langle C_1\rangle  - \langle A_1C_1\rangle  - \langle B_1C_1\rangle  - \langle A_1B_1C_1\rangle ~\leq~1$ \\[.2em] \hline
2 & $ - 2\langle A_1\rangle  - 2\langle B_1\rangle  - \langle C_0\rangle  - \langle A_0C_0\rangle  - \langle B_0C_0\rangle  - \langle A_1B_0C_0\rangle  - \langle A_0B_1C_0\rangle  + \langle A_1B_1C_0\rangle  - \langle C_1\rangle  + \langle A_0C_1\rangle  + \langle B_0C_1\rangle  + \langle A_1B_0C_1\rangle  + \langle A_0B_1C_1\rangle  + \langle A_1B_1C_1\rangle ~\leq~4$ \\[.2em] \hline
3 & $ - \langle A_0\rangle  - \langle A_1\rangle  - \langle A_0B_0\rangle  + \langle A_1B_0\rangle  - 2\langle B_1\rangle  - \langle C_0\rangle  - \langle A_0C_0\rangle  + \langle B_0C_0\rangle  + \langle A_1B_0C_0\rangle  - \langle A_0B_1C_0\rangle  + \langle A_1B_1C_0\rangle  - \langle C_1\rangle  + \langle A_1C_1\rangle  - \langle B_0C_1\rangle  - \langle A_0B_0C_1\rangle  + \langle A_0B_1C_1\rangle  + \langle A_1B_1C_1\rangle ~\leq~4$ \\[.2em] \hline
4 & $ - 2\langle A_1\rangle  + 2\langle A_0B_0\rangle  - \langle C_0\rangle  - \langle A_0C_0\rangle  + \langle A_0B_0C_0\rangle  - \langle A_1B_0C_0\rangle  + \langle B_1C_0\rangle  + \langle A_1B_1C_0\rangle  - \langle C_1\rangle  + \langle A_0C_1\rangle  + \langle A_0B_0C_1\rangle  + \langle A_1B_0C_1\rangle  + \langle B_1C_1\rangle  + \langle A_1B_1C_1\rangle ~\leq~4$ \\[.2em] \hline
5 & $ - \langle A_0\rangle  + 2\langle A_0B_0\rangle  - \langle B_1\rangle  + \langle A_0B_1\rangle  - \langle C_0\rangle  - \langle A_0C_0\rangle  + \langle A_0B_0C_0\rangle  - \langle A_1B_0C_0\rangle  + \langle B_1C_0\rangle  + \langle A_1B_1C_0\rangle  - 2\langle C_1\rangle  + 2\langle A_1C_1\rangle  + \langle A_0B_0C_1\rangle  + \langle A_1B_0C_1\rangle  + \langle A_0B_1C_1\rangle  + \langle A_1B_1C_1\rangle ~\leq~5$ \\
\hline

\end{longtable*}

\begin{longtable*}{|c|c|c|c|c|}
\caption{$NS_2$, $T_2$ and $S_2$ bounds for the inequalities of Table~\ref{tvertphys}, as well as maximum quantum value.\\
$^*$ Best value found by considering three-qubits states.\\
$^\dag$ Best upperbound found with the NPA hierarchy.}\label{tbounds}\\
\hline
Class & $NS_2$ bound & $T_2$ bound & $S_2$ bound & quantum bound \\
\hline
\endfirsthead
\caption{continued}\\
\hline
\ Class\ &\ $NS_2$ bound\ &\ $T_2$ bound\ &\ $S_2$ bound\ &\ quantum maximum\ \\
\hline
\endhead
  \multicolumn{5}{l}{{Continued on the next page...}}
\endfoot
  \hline
\endlastfoot
1 & 1 & 1 & 1 & 1.00000\\
\hline
2 & 4 & 4 & $32/5= 6.4$  & 4.75069\\
\hline
3 & 4 & 6 & $20/3\simeq 6.66667$  & 4.81193\\
\hline
4 & 4 & 6 & 7 & 5.00000\\
\hline
5 & 5 & 7 & $23/3\simeq 7.66667$  & 5.50921\\
\hline
6 & 7 & 7 & $73/7\simeq 10.42857$  & 8.12659\\
\hline
7 & 7 & 9 & $51/5= 10.2$  & 7.95393\\
\hline
8 & 6 & 8 & 9 & 6.56259\\
\end{longtable*}

\end{document}